\documentclass[twocolumn,showpacs,amssymb,amsmath,floatfix,superscriptaddress]{revtex4}
\usepackage{graphicx,amssymb}

\begin{document}
\title{Lifetimes of electrons in the Shockley surface state band of Ag(111)}

\author{M. Becker}
\email{becker@physik.uni-kiel.de}
\affiliation{Institut f\"ur Experimentelle und Angewandte Physik, 
Christian-Albrechts-Universit\"at
zu Kiel, D-24098 Kiel, Germany}

\author{S. Crampin}
\affiliation{Department of Physics, University of Bath, Bath BA2 7AY, 
United Kingdom}

\author{R. Berndt}
\affiliation{Institut f\"ur Experimentelle und Angewandte Physik, 
Christian-Albrechts-Universit\"at
zu Kiel, D-24098 Kiel, Germany}

\date{31.08.2006}

\begin{abstract}
We present a theoretical many-body analysis of the electron-electron ({\it e-e}) inelastic damping rate $\Gamma$ of electron-like excitations in the Shockley surface state band of Ag(111). 
It takes into account {\it ab-initio} band structures for both bulk and surface states. $\Gamma$ is found to increase more rapidly as a function of surface state energy $E$ than previously reported, thus leading to an improved agreement with experimental data. 
\end{abstract}
\maketitle

\section{Introduction}
The quantitative understanding of the dynamics of electronic excitations in the Shockley state band at the (111) face of Ag has been the subject of a considerable number of research reports (for a review see Ref.\ \cite{ech04_}).
Recent experimental studies used
photoelectron spectroscopy and the scanning tunnelling microscope (STM) \cite{rei01_,jli98_,bur99_,kli00a,kli01_,bra02_,vit03_,jen05_}.
  
Theoretically, many-body lifetime calculations using the GW approximation \cite{kli00a,lek03_} have clarified the role of electron-electron ({\it e-e}) interactions as well as the relative importance of intraband and interband scattering and screening. The damping rate due to electron-phonon scattering is also quantitatively understood \cite{eig03_}. However, comparing theoretical with STM-derived experimental damping rates of electronic excitations with energies above the Fermi level $E_F$ show that the experimental decay rates increase more rapidly with increasing energy $E$.
Recently a similar discrepancy between 
image state lifetimes measured by
means of the STM and two-photon-photoemission
spectroscopy was lifted \cite{cra05c}. It was shown that the
applied electric field during the STM-measurement causes an increase in
both the efficiency of the image state decay channels as well as their
number. In the light of this work, Becker {\it et al.}  theoretically investigated the impact of the STM-induced electric field on the inelastic {\it e-e}
damping rate of the electron- and hole-like excitations in the Shockley
surface state band at Ag(111) \cite{bec06a}. Their results indicate that under typical tunnelling conditions the STM does not significantly alter the surface
state wave function and so previous STM-derived Shockley state lifetimes
need not be corrected. Nevertheless, while experimental and calculated lifetimes agree quite well over a range of energies, a quantitative difference remains for unoccupied states of energies $E\gtrsim E_F+300$~meV. The origin of this different energy dependence, which is still an open question, is addressed in this report.

\section{Calculation details}

The calculations presented here are based upon the approach developed by Chulkov {\it et al.} \cite{chu98_}, and used widely in calculations of surface 
state dynamics \cite{ech04_}. 
Within this approach the damping rate or inverse lifetime of an excitation in the
state $\psi(\mathbf{r})$ with energy $E$ is obtained from the 
expectation value of the imaginary part of the electron self-energy, 
$\Sigma(\mathbf{r},\mathbf{r}';E)$:
\begin{equation}
\Gamma=\tau^{-1}=
-2\int\!\! d\mathbf{r}\!\int \!\!d\mathbf{r}' \, \psi^{*}(\mathbf{r})\mathrm{Im}
\Sigma(\mathbf{r},\mathbf{r}';E)\psi(\mathbf{r}').
\label{eqn:gamma}
\end{equation}
Unless stated explicitly, atomic units are used throughout the text, i.e., $e^2=\hbar=m_e=1$.

\noindent
In the GW approximation \cite{hed69_} of many-body theory the imaginary
part of the self energy is calculated to first order in terms of the screened Coulomb
interaction $W$ and the Green function $G$
\begin{equation}
\mathrm{Im}\Sigma(\mathbf{r},\mathbf{r}';\epsilon)\!=\!
-\frac{1}{\pi}\int_{E_F}^\epsilon \!\!d\epsilon'\,
\mathrm{Im}G(\mathbf{r},{\mathbf r}';\epsilon')
\mathrm{Im}W(\mathbf{r},\mathbf{r}';\epsilon\!-\!\epsilon').
\label{eqn:imsig}
\end{equation}
Here we apply the widely used zeroth order approximation 
to the Green function in the spectral representation
\begin{equation}
G(\mathbf{r},\mathbf{r}';\epsilon)=
\sum_{i}\frac{\psi_{i}(\mathbf{r})\psi_{i}^{*}(\mathbf{r}')}
{\epsilon-E_{i}+{\rm i}\delta},
\label{eqn:g}
\end{equation}
where the $\psi_{i}(\mathbf{r})$ are one-electron eigenfunctions with 
energies $E_{i}$ and $\delta$ a positive infinitesimal. The screened Coulomb interaction is evaluated in the random phase 
approximation (RPA)
\begin{eqnarray}
W(\mathbf{r},\mathbf{r}';\omega)&=&v(\mathbf{r}-\mathbf{r}')+\int d\mathbf{r}_1\int d\mathbf{r}_2
v(\mathbf{r}-\mathbf{r}_1)\nonumber\\
&&\times \chi^{0}(\mathbf{r}_1,\mathbf{r}_2;\omega)W(\mathbf{r}_2,\mathbf{r}';\omega)
\label{eqn:w}
\end{eqnarray}
where $v$ is the bare Coulomb interaction and $\chi^{0}$
is the density-density response function of the non-interacting electron
system. An explicit expression for $\chi^0$ can be found in Ref.\ \cite{egu83}.

The single particle states have been obtained
by solving the Schr\"odinger equation using a one-dimensional
pseudopotential varying in the direction perpendicular to the surface
\cite{chu99_}.
The pseudopotential does not describe the $d$-electrons of the substrate,
which are anyways too low in energy to play a significant
role as final states for the decay,
but their contribution to the screening is included via
a polarisable background \cite{lek03_,lie93_}.
\begin{figure}[t!]
\includegraphics[height=65mm, width=70mm]
{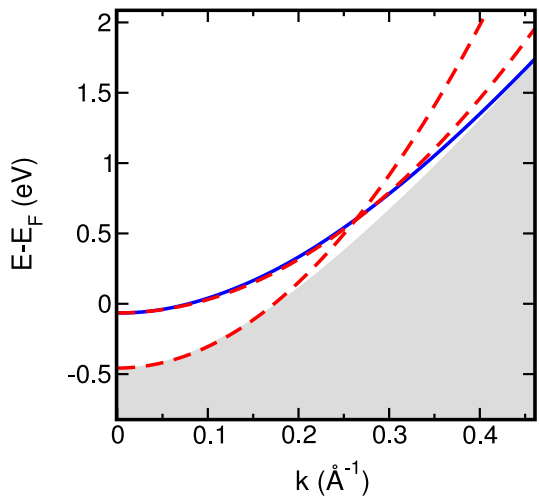}
\caption{
\label{fig:disp}
(Color online) Ag(111) surface band structure. Shaded area and solid line are the
projected bulk and surface state dispersion from {\it ab-initio} 
calculations which we use for calculating the {\it e-e} decay rate.
The dashed lines are parabolic dispersions with effective masses $m^*=0.4$
(surface state band) and $0.25$ (band edge).}
\end{figure}
The electron-phonon contribution to the linewidth is taken to be constant $\Gamma_{\rm e-ph}=3.7$~meV for energies in excess of $20$~meV above the Fermi level \cite{eig03_}.

Previous calculations of the decay rate of the Ag(111) Shockley state
have assumed parabolic dispersion with effective masses of $m^*=0.4$
and $m^*=0.25$ for the intrinsic surface state and
the lower band edge respectively \cite{vit03_}.
However, over the extended energy
range of interest here, this approximation needs to be improved as evident from Fig.\ \ref{fig:disp}. 
Figure \ref{fig:disp} shows the parabolic dispersions (dashed lines) for the surface state band and the band edge with effective masses as mentioned above. The grey shaded area and the solid line are the projected bulk and surface state dispersion, respectively, that are found from {\it ab-initio} calculations. The parabolic dispersion underestimates the gap between surface state band and band edge. We have therefore
used the {\it ab-initio} dispersions for our lifetime calculations.
As previously noted \cite{vit03_,ver05_} there are important changes in the 
shape of the surface state wave function with $k$, the surface wave vector,
and we take these into account by recalculating the wave function
for different $k$ with the pseudopotential parameters changed to take into
account the appropriate {\it ab-initio} band edges and surface state energy.

\section{Results and discussion}

\begin{figure}[t!]
\includegraphics[
height=65mm, width=70mm]
{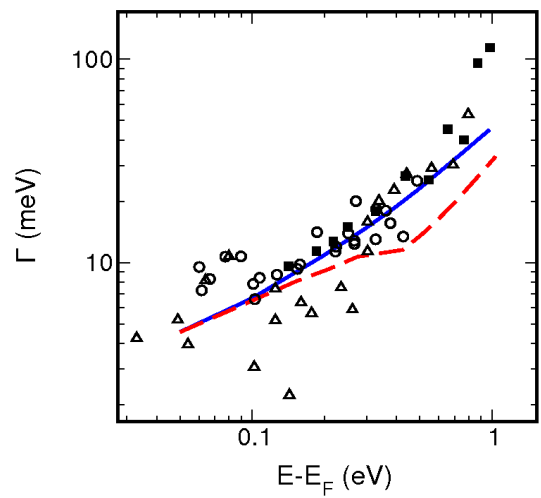}
\caption{
\label{fig:damp}
(Color online) Comparison of the total Shockley surface state decay rates $\Gamma+\Gamma_{\rm e-ph}$ for electrons at Ag(111) calculated using {\it ab-initio} dispersion relations (solid line) and previous published decay rates using parabolic dispersion (dashed line) from Ref.\,\cite{ech04_}.
The black squares represent experimental data from Ref. \cite{vit03_}, black triangles from Ref.\,\cite{bra02_} and black circles from Ref.\,\cite{jen05_}. The experimental data from Refs.\,\cite{bra02_,vit03_} have been multiplied by a factor 2 to compensate for an erroneous analysis in that work \cite{cra05a}. 
}
\end{figure}

We focus on the energy range between $0$ and $1$~eV above the Fermi level where the surface state is energetically
separated from the bulk states and thus is well defined. This energy range is of interest as according to Ref.\ 
\cite{vit03_} there is a 
cross-over in the primary decay channel from intraband to interband dominated 
decay.
Our calculated results for the total damping rate $\Gamma+\Gamma_{\rm e-ph}$ are shown in Fig. \ref{fig:damp} as a solid line.  
For energies below $E-E_F\approx 0.1$~eV we find that our results for the inelastic {\it e-e} damping rate of excitations in the Shockley state band agree well with the inverse lifetime estimation from Echenique {\it et al.} (dashed line) \cite{ech04_} using similar methods but with an effective mass approximation. 
This agreement is due to the fact that the effective mass approximation gives a very good description 
of the projected {\it ab-inito} surface band structure for energies with $E-E_{F}\lesssim0.1$~eV as evident from Fig.~\ref{fig:disp}. 

For larger energies we find that the
{\it e-e} damping rate as a function of energy increases more rapidly with the {\it ab-initio} dispersion than calculated with parabolic dispersions.  This result can be understood as follows: 
For energies $E-E_F\gtrsim0.1$~eV 
the parabolic dispersions underestimate the energy gap between surface state band and the bulk band edge.
Consequently, the {\it ab-initio} dispersions reveal a reduced penetration of the surface state wave function into the
interior of the crystal and an increased probability amplitude in the proximity of the metal surface. 
This modification leads to two competing effects. Since screening is reduced at the surface \cite{kli00a} 
intraband decay becomes more efficient. On the other hand, interband decay becomes less probable owing to the reduced overlap with bulk states.  It turns out that the increase of intraband decay outweighs the decrease in interband transitions.
The net result is a more steep increase of $\Gamma$   than obtained from parabolic dispersions.
We thus find the intraband contribution to dominate up to $E-E_F=1$~eV where the interband and intraband contributions become similar in magnitude.
Comparing to published experimental data (symbols in Fig.\ \ref{fig:damp}) the agreement is improved.

In summary, we have presented a  theoretical many-body analysis of the inelastic {\it e-e} lifetimes of electronic excitations in the Shockley surface state band of Ag(111). The calculations are based on the GW approximation  using a one-dimensional pseudopotential together with an {\it ab-initio} description of the projected surface band structure. We have shown that the consistency of experimental and theoretical derived damping rates is significantly improved when using {\it ab-initio} band structures instead of an effective mass approximation for calculating the decay rate.

\section{Acknowledgements}
M.B. and R.B. thank the Deutsche Forschungsgemeinschaft for financial support through {\it SPP 1093}.

\end{document}